\documentclass[compsoc,conference,a4paper,10pt,times]{IEEEtran}
\IEEEoverridecommandlockouts
\usepackage{cite}
\usepackage{amsmath,amssymb,amsfonts}
\usepackage{algorithmic}
\usepackage{graphicx}
\usepackage{textcomp}
\usepackage{bmpsize}
\usepackage[table,xcdraw]{xcolor}
\usepackage{lipsum}
\usepackage{balance}
\usepackage{stfloats}
\usepackage[colorlinks=true,urlcolor=black]{hyperref}
\def\BibTeX{{\rm B\kern-.05em{\sc i\kern-.025em b}\kern-.08em
    T\kern-.1667em\lower.7ex\hbox{E}\kern-.125emX}}
\usepackage{newtxtext,newtxmath}

\newif\ifcomments

\usepackage{xspace}                         %
\usepackage{caption}
\usepackage{subcaption}
\usepackage[normalem]{ulem}                 %
\usepackage{upgreek}
\usepackage{comment}                        %
\usepackage{rotating}
\usepackage{graphicx}
\usepackage{multirow}
\usepackage{multicol}
\usepackage{cleveref}
\usepackage{listings,lipsum}
\usepackage{gensymb}  %
\usepackage[htt]{hyphenat}  %
\usepackage{tikz}
\usepackage{ctable}
\usepackage{xurl}
\usepackage[table,xcdraw]{xcolor}

\usepackage[skip=0pt]{caption}
  \usepackage{ifthen}

\makeatletter
\@ifclassloaded{sigplanconf}{
}{
}
\makeatletter

\newcommand{\printlayout}{{\bf The textwidth \the\textwidth; the columnwidth is \the\columnwidth}}

\newcommand{\td}[1]{\texttt{SD#1}}

\newcommand{\eg}{\textit{e.g.,}\xspace}
\newcommand{\ie}{\textit{i.e.,}\xspace}

\newcommand{\microsecond}{$\upmu{}$s\xspace}

\newcommand{\amd}{\textsf{x86\_64}}
\newcommand{\arm}{\textsf{Arm}}
\newcommand{\riscv}{\textsf{RISC-V}}

\newcommand{\vtx}{\textsf{VT-x}}
\newcommand{\iommu}{\textsf{I/O-MMU}}

\newcommand{\mmode}{\textsf{M}-mode}
\newcommand{\hmode}{\textsf{H}-mode}

\newcommand{\manifest}{boot-info}

\newcommand{\nbloccapa}{4K}
\newcommand{\nbruntimebytes}{230KB}

\newcommand{\cpuamd}{Intel i7-10700}
\newcommand{\cpuriscv}{StarFive VisionFive2 board}
\newcommand{\linuxamd}{v6.2}
\newcommand{\linuxriscv} {v5.15}

\newcommand{\riscvtrans}{$3.9$\microsecond}
\newcommand{\amdtrans}{$1.2$\microsecond}

\newcommand{\myparagraph}[1]{\vspace{0.2em}\noindent {\bf #1:}}

\lstnewenvironment{attestation}[1][]{%
  \lstset{
    basicstyle=\ttfamily\linespread{0.7}\small,
    frame=tb,
    emph={alias, aliased, carve, carved, root, domain, with, signature, for, exclusive},
    emphstyle=\textbf
    #1
  }%
}{}

\lstnewenvironment{monitor}[1][]{%
  \lstset{
    language=C, 
    basicstyle=\ttfamily\small,
    frame=tb,
    emph={create, alias, carve, send, seal, switch, enumerate, attest, get, set},
    emphstyle=\textbf,
    numbers=left, stepnumber=1, numbersep=5pt,
    commentstyle=\color{gray},
    escapeinside={(*@}{@*)},
    #1
  }%
}{}

\usepackage[english]{babel}
\addto\extrasenglish{%
  \def\chapterautorefname{\S\@gobble}%
  \def\sectionautorefname{\S\@gobble}%
  \def\subsectionautorefname{\S\@gobble}%
  \def\subsubsectionautorefname{\S\@gobble}%

  }

\newcommand{\system}{\textsc{Tyche}\xspace}
\newcommand{\framework}{\system{} SDK}

\newcommand{\capadriver}{\system{}-Capa\xspace}
\newcommand{\kvmdriver}{KVM-\system{}\xspace}
\newcommand{\kvmintel}{KVM-Intel}

\newcommand{\graminetyche}{Gramine-\system{}\xspace}

\newcommand{\cveswitchlessone}{CVE-2022-21233}
\newcommand{\cveswitchlesstwo}{CVE-2022-21166}

\newcounter{designcounter} %

\newcommand{\epcsize}{94MB}
\newcommand{\perfllama}{$\sim$2\%}
\newcommand{\llamanbtokens}{1000}
\newcommand{\llamamodel}{Meta LLaMa 3.2 Instruct model}
\newcommand{\rvslowdownqsort}{\textless{}10\%}
\newcommand{\tdtwothreadslowdown}{$\sim${}2\%}
\newcommand{\tdvmcpuoverheadnativevm}{$\sim${}1\%}
\newcommand{\tdvmcpuoverheadnative}{$\sim${}4\%}

\newcommand{\speedtestsize}{1000}
\newcommand{\speedtestinserts}{500K}

\begin{document}

\title{Tyche: Composable Isolation as a Foundation to Manage Trust in the Cloud}

\makeatletter
\newcommand{\linebreakand}{%
  \end{@IEEEauthorhalign}
  \hfill\mbox{}\par
  \mbox{}\hfill\begin{@IEEEauthorhalign}
}

\author{\IEEEauthorblockN{Adrien Ghosn\textsuperscript{*}\thanks{* Co-equal first author.}}
\IEEEauthorblockA{\textit{Azure Research, Microsoft} \\
\textit{Cambridge, UK}\\
adrienghosn@microsoft.com}
\and
\IEEEauthorblockN{Charly Castes\textsuperscript{*}}
\IEEEauthorblockA{\textit{EPFL} \\
\textit{Lausanne, Switzerland}\\
charly.castes@epfl.ch}
\and
\IEEEauthorblockN{Neelu S. Kalani}
\IEEEauthorblockA{\textit{EPFL} \\
\textit{Lausanne, Switzerland}\\
neelu.kalani@epfl.ch}
\linebreakand
\IEEEauthorblockN{Yuchen Qian}
\IEEEauthorblockA{\textit{EPFL} \\
\textit{Lausanne, Switzerland}\\
yuchen.qian@epfl.ch}
\and
\IEEEauthorblockN{Marios Kogias}
\IEEEauthorblockA{\textit{Imperial College London} \\
\textit{London, UK}\\
m.kogias@imperial.ac.uk}
\and
\IEEEauthorblockN{Edouard Bugnion}
\IEEEauthorblockA{\textit{EPFL} \\
\textit{Lausanne, Switzerland}\\
edouard.bugnion@epfl.ch}
}

\maketitle

\begin{abstract}

Cloud workloads combine software components from different parties to process sensitive data.
Each component has its own trust model---it must protect its assets from the rest of the system, yet share sensitive data with components it cannot trust to keep confidential.
This tension requires composing isolation boundaries for confidentiality and encapsulation.
Unfortunately, the cloud offers no direct way to compose such boundaries, forcing tenants to assemble, deploy, and maintain their own solutions.
This paper shifts that burden back to the infrastructure by making composable, attestable isolation a first-class systems abstraction.

We present \system{}, a security monitor that centers isolation around a unified composable abstraction: security domains (\td{}s).
An \td{} is an execution environment whose access to machine resources---memory, cores, devices---is controlled through explicit capabilities.
A small set of capability operations enables \td{}s to partition, share, and reclaim resources; by nesting recursively, \td{}s compose attestable trust boundaries for confidentiality and encapsulation.
\system{} attests these compositions, providing end-to-end security guarantees for workloads made of mutually distrustful components.
As a first-class cloud primitive, this single abstraction subsumes enclaves, sandboxes, CVMs, and their compositions.

\system{} provides composable isolation without sacrificing compatibility with existing hardware and software stacks.
It runs on commodity \amd{} hardware without security extensions, and a \riscv{} prototype demonstrates portability across platforms.
Our SDK composes isolation for unmodified workloads within \td{}s with minimal overhead.
In a confidential LLM inference scenario with mutually distrustful users, model owners, and cloud providers, the slowdown is just 2\% compared to bare-metal Linux.

\end{abstract}

\section{Introduction}

The cloud lacks abstractions for managing heterogeneous trust.
Modern workloads are multi-party: even within a single virtual machine (VM), container, or process,
components supplied by independent parties -- libraries, frameworks, services, CSP infrastructure -- must cooperate to process sensitive data, but none fully trusts the others.
A component must protect its own assets, such as ML models or proprietary code, while handing sensitive data to untrusted components that perform essential processing but may leak it.
This simultaneous need for confidentiality and encapsulation is what we call \textit{controlled sharing}.
It is inherent to cloud deployments, yet no cloud abstraction directly supports it.

Cloud abstractions, \eg containers and VMs, ease deployment but offer only coarse-grained trust management.
Even trusted execution environments (TEEs) are limited to a binary trust model: software inside an enclave~\cite{sgx, trustzone} or a confidential VM (CVM)~\cite{tdx, amd-sev-snp-patch, cca} is trusted, and its confidentiality and integrity is guaranteed against privileged code, \eg the CSP's hypervisor.
Yet managing heterogeneous trust within the TEE itself, and enforcing controlled sharing among its components, is left entirely to the tenant.

Tenants address this burden in one of two ways: (1) external isolation, running components in separate TEEs, or (2) intra-isolation, inside a single TEE.

In the external approach, each component runs in its own CSP-provided TEE.
The TEEs mutually attest before exchanging encrypted sensitive data over untrusted memory.
This protects the confidentiality of each TEE against other components, but fails to guarantee encapsulation once one TEE hands data to another.
Preventing unintended leakage thus requires trusting or verifying the functional correctness of each TEE---an unrealistic assumption when TEEs contain proprietary code, a full commodity operating system and applications, or require frequent updates.

\begin{figure}
    \includegraphics[width=\columnwidth]{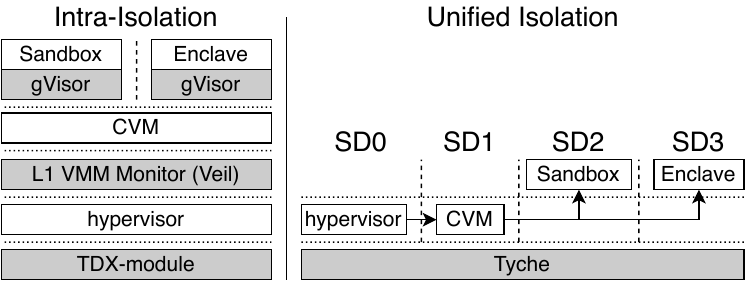}
    \caption{
        Composing enclaves and sandboxes within a CVM. Left: Intel TDX intra-isolation, with gVisor~\cite{gvisor} sandboxing both an application (sandbox) and a VMM-provided enclave. Right: our approach. Arrows indicate management dependencies.
    }
    \label{fig:trust-domains}
\end{figure}

Alternatively, intra-isolation composes isolation by creating boundaries within the TEE to separate its software components.
By leveraging the TEE's internal privilege hierarchies---rings, AMD SEV-SNP's VMPLs~\cite{sev-snp}, TDX L1/L2 partitions~\cite{tdx, tdx-partitioning}---privileged software can enforce isolation for less-privileged components.
This enables nested enclaves for confidentiality~\cite{veil, verismo}, protecting assets from the rest of the CVM's code, or sandboxes for encapsulation~\cite{erebor, gvisor}, preventing untrusted components from leaking sensitive data.

Intra-isolation is a popular research direction, and \autoref{tab:comparison-matrix} surveys representative examples.
However, the space is fragmented: each system provides nested isolation for a fixed threat model and platform, and none are designed to interoperate.
Existing systems target different threat models, nested sandboxes~\cite{erebor} or enclaves~\cite{veil} on different platforms (\eg Veil~\cite{veil} on AMD, Erebor~\cite{erebor} on Intel) or compete for the same privilege layer (\eg Veil and Verismo~\cite{verismo} both require VMPL0), making these systems hard to compose.
Hardware further limits composition, capping the number of sub-components (\eg only 1+3 partitions in TDX~\cite{tdx-partitioning}) or prohibiting certain combinations outright (\eg no SGX enclaves inside TDX CVMs).

Worse, even setting these incompatibilities aside, combining separate solutions in a single deployment compounds complexity and is counter-productive for security.
Consider private LLM inference: the user must reveal prompts to an untrusted LLM runtime without allowing exfiltration, while the model owner must keep proprietary weights hidden from both user and CSP.
This demands composing nested enclaves and sandboxes within a single CVM, a typical case of controlled sharing.
In \autoref{fig:trust-domains}(left), an L1 VMM monitor (\eg similar to Veil~\cite{veil}) provides a nested enclave to protect the LLM runtime and model weights, while the user controls the surrounding CVM and uses gVisor~\cite{gvisor} to encapsulate the enclave and prevent prompt leakage or sandbox other applications.
Composing isolation further, \eg sandboxing libraries inside the enclave, deepens the hierarchy and would require Wasm~\cite{wasm} or NaCl~\cite{nacl} as in Ryoan~\cite{ryoan}.
Each layer duplicates enforcement, expands the trusted computing base (TCB), and forces tenants to integrate, configure, and maintain complex sub-systems.
Correct isolation becomes fragile and error-prone, and attesting end-to-end security grows complicated by uncertainty about which layers must be measured.
Here, to ensure prompts are not leaked, the user must attest the TDX module, the CVM, and gVisor, but also the monitor to ensure it does not provide enclaves with a covert channel to bypass the gVisor sandbox.

\begin{table*}[]
	\center
	\footnotesize
	\begin{tabular}{|l|l|l|l|c|c|}
		\hline
		\textbf{Outer \textbackslash{} Inner} & \multicolumn{1}{c|}{\textbf{No Nesting}}               & \multicolumn{1}{c|}{\textbf{Sandbox}} & \multicolumn{1}{c|}{\textbf{Enclave}}          & \textbf{VM}                                 & \textbf{CVM}                                   \\ \hline
		\textbf{Sandbox}                      & gVisor~\cite{gvisor}, Wasm~\cite{wasm}                 & Capsicum~\cite{capsicum}              & \multicolumn{1}{c|}{\cellcolor[HTML]{C0C0C0}-} & \cellcolor[HTML]{C0C0C0}-                   & \cellcolor[HTML]{C0C0C0}-                      \\ \hline
		\textbf{Enclave}                      & SGX~\cite{sgx}, Nitro Enclave~\cite{aws-nitro-enclave} & Ryoan~\cite{ryoan}                    & Nested-Enclave~\cite{nested-enclaves}          & \cellcolor[HTML]{C0C0C0}-                   & \cellcolor[HTML]{C0C0C0}-                      \\ \hline
		\textbf{VM}                           & VT-x~\cite{vtx}                                        & Hyperlight~\cite{hyperlight}          & SGX~\cite{sgx}                                 & \multicolumn{1}{l|}{KVM Nested-Virt}        & \multicolumn{1}{l|}{Hyper-V confidential L2}   \\ \hline
		\textbf{CVM}                          & TDX~\cite{tdx}, SEV-SNP~\cite{sev-snp}                 & Erebor~\cite{erebor}                  & Veil~\cite{veil}                               & \multicolumn{1}{l|}{OpenHCL~\cite{openhcl}} & \multicolumn{1}{c|}{\cellcolor[HTML]{C0C0C0}-} \\ \hline
	\end{tabular}
	\caption{
		Nested composition of isolation abstractions in prior work.
		Rows denote the outer isolation boundary and columns the inner boundary.
		Prior systems support only specific combinations, each requiring dedicated hardware or software extensions, motivating the need for a unified composable isolation solution.
	}
	\label{tab:comparison-matrix}
\end{table*}

Our insight is that intra-isolation complexity is a symptom, not a necessity: it reflects the lack of a composable isolation abstraction in the cloud.
All isolation mechanisms---regardless of trust model or platform---ultimately enforce access and boundary control, and differ only in where and how they apply it.
Intra-isolation stacks them because the cloud offers no way to compose access and boundary control directly.
A single enforcement layer that natively supports composition would eliminate the duplication entirely.

We thus propose a third approach that makes composable isolation a first-class cloud abstraction.
A single trusted entity provides and attests composable access and boundary control.
By making composition a core feature, this entity allows components to compose attestable isolation boundaries for diverse trust models, without duplicating enforcement.
The result is lower deployment complexity, a smaller software and hardware TCB, and precise management of nuanced trust relationships in the cloud.

Our approach raises two key research questions:
(1) \textit{What primitives enable composable isolation for diverse trust models?}
(2) \textit{How can we introduce composable isolation without breaking existing software and hardware stacks?}
These questions guide our design toward a general yet practical solution.

We realize this approach in \system{}, a security monitor that provides composable isolation through a unified abstraction: security domains (\td{}s).
An \td{} is an execution environment whose access to machine resources is controlled through explicit capabilities.
A small set of capability operations---creating \td{}s, partitioning, sharing, and reclaiming resources---lets \td{}s \textit{recursively compose attestable isolation boundaries}.
\system{} attests these compositions, making access and boundary control explicit end-to-end.
This subsumes prior mechanisms: exclusive access supports confidentiality and integrity, while control over shared resources and interactions enables encapsulation.

\system{} runs on bare metal at the highest privilege level, attested by a hardware root of trust~\cite{tpm}, and enforces isolation for all other software running as \td{}s.
Implemented in Rust~\cite{rust}, it uses standard hardware access-control mechanisms and supports deployment across platforms.
We present a full implementation on \amd{} using virtualization technologies~\cite{vtx, vtd} and a firmware prototype on \riscv{}~\cite{pmp}.
For compatibility, our \framework{} allows Linux \td{}s to create and compose sandboxes, enclaves, and CVMs as nested \td{}s.
This replaces ad-hoc intra-isolation mechanisms (\autoref{fig:trust-domains}, right) and covers the entire design space of \autoref{tab:comparison-matrix} with unified semantics, attestation, and enforcement.

Our evaluation on \amd{} shows that \system{} \td{}s---replacing enclaves, VMs, CVMs, and nested enclaves within CVMs---perform comparably to their native equivalents across diverse workloads, including web servers, databases, key-value stores, and LLM inference.
In a confidential LLM inference scenario with mutually distrustful users, model owners, and CSPs, the slowdown is just 2\% compared to bare-metal Linux.
\system{} is an open-source research prototype~\cite{anon} that independent researchers used to achieve custom isolation guarantees on legacy hardware~\cite{argos}.

The remainder of this paper covers background and motivation (\autoref{sec:motivation}), a high-level overview (\autoref{sec:overview}), \system{}'s capability API (\autoref{sec:capabilities}), its platform-specific implementation (\autoref{sec:implementation}), and the \framework{} (\autoref{sec:deploy}).
We then evaluate \system{} (\autoref{sec:evaluation}) and discuss inspiration drawn from related work in monitors and micro-kernels (\autoref{sec:related-work}).

\section{Motivation: See the forest for the T(r)EEs}
\label{sec:motivation}

\myparagraph{Observations}
Computer system security requires resource (\ie memory, CPU, interrupts, devices) and fault isolation between software components that implement different tasks or services.
Isolation is however rarely absolute, as components need to cooperate to carry out computations.
For instance, two processes might share memory; isolation between a VM and a hypervisor is unidirectional to enable, \eg emulated devices or migration.
Isolation in systems is thus more accurately described as an attempt to enforce the least privilege principle~\cite{segmentation}: each component should be granted only the minimal access rights it needs to perform its intended operation.
Historically, systems were designed under the assumption that a single organization controlled the entire software stack and hardware, but that assumption no longer holds as cloud platforms and modern deployments combine components from multiple independent sources and place critical privileged components under the control of external providers in the cloud.

As modern systems integrate components from diverse sources, enforcing least privilege increasingly relies on \textit{compartmentalization}: structuring the system so that each component's access to the machine is tightly bounded.
\textit{Confidential computing} provides one form of compartmentalization by preventing access from more privileged code.
It departs from traditional system designs that conflate privilege with trust and treat resource management as justification for access.
Examples include enclaves~\cite{komodo, trustvisor, sgx} and CVMs~\cite{cca, tdx, sev-snp}, which differ mainly in how they interface with the surrounding system, not in their underlying trust model.
Another form of compartmentalization is \textit{encapsulation}, which restricts how a component may interact with the rest of the system, often to prevent information leaks or interference with other components.
Sandboxing mechanisms~\cite{nacl, wasm, lfi}, as well as containers, are examples of encapsulation.

While often treated separately, confidential computing and encapsulation are two sides of the same coin~\cite{tyche}.
Both aim at controlling access to resources.
Encapsulation focuses on \textit{sharing}: explicitly granting a component-controlled access to resources such as memory or interfaces.
Confidential computing, in contrast, focuses on \textit{exclusivity}: ensuring that a component's resources cannot be accessed by others, either via access control mechanisms~\cite{sgx, trustvisor, blackbox} or encryption and integrity guarantees~\cite{tdx, sev}.
Existing solutions often emphasize only one side, reflecting the trust model of a particular component and producing rigid abstractions -- for example, the binary notion of trust in TEEs -- without recognizing the common underlying goal.
From a holistic cloud security perspective, all solutions are mere instances of access and boundary control.
Such control enables higher-level properties such as confidentiality, integrity, encapsulation, and controlled sharing: the ability to selectively expose component resources while guaranteeing they cannot be leaked to unintended parties.
To address the varying components' trust models in a coherent whole, isolation needs to be \textit{composed}, \ie components must be able to define and control their own boundaries.

We define composable isolation as the ability to assemble isolation boundaries in a way that preserves and combines their security guarantees.
Each isolation boundary can be seen as a function that refines the access and boundary control of the components it contains: it restricts what they can access and with whom they can interact.
Composing isolation is then composing these functions: each new boundary refines the previous one, analogous to function composition in mathematics where $f \circ g(x) = f(g(x))$: each layer further tightens access and boundary control.
This composition can be enforced by hierarchical systems that each apply one refinement---first $g$, then $f$---as intra-isolation does today, or by a single layer that enforces the composed projection $f \circ g$ directly.

In a cloud setting, where control over software and hardware may lie with an external party, isolation must be \textit{attested}.
Attestation demonstrates to a local or remote entity that a known software version is running on a platform and is isolated from the rest of the system according to the intended security policies.
It typically relies on a hardware root of trust, such as the CPU~\cite{sgx, tdx, sev} or a Trusted Platform Module (TPM)~\cite{tpm} and may also involve trusted software.

\myparagraph{Lessons learned and goals}
From the discussion above, two key properties emerge as fundamental for composing isolation in heterogeneous cloud systems.
First, management must be decoupled from access, and trust decoupled from privilege, reflecting the insight from TEEs.
Second, access and boundary control must be attestable, so it is possible to verify that a component's interactions and resource accesses respect intended policies.
These lessons address our first research question:
\textit{What primitives enable composable isolation for diverse trust models?}
We argue that the two sufficient primitives are \textit{attestation}, and \textit{access and boundary control}.
With those, we can not only re-construct existing higher-level abstractions but also provide finer-grained and composable isolation.

\myparagraph{Approach}
Providing these primitives while addressing our second research question, \ie preserving the existing software stack, is challenging and requires careful design and implementation.
In particular, it requires designing an interface that is small and easy to integrate with existing software, while implementing an enforcement that does not disrupt existing software hierarchies.
For this, we identify capabilities and security monitors as attractive candidates.

Capabilities provide a structured and explicit mechanism for access control.
They are unforgeable tokens -- created either in software by a trusted entity~\cite{sel4} or implemented in hardware~\cite{cheri, cheriot} -- that encapsulate access to an object.
Holding a capability grants specific access rights to the object, while transferring it to another component revokes access from the sender and grants it to the recipient.
Capabilities are used by security-focused kernels~\cite{sel4} and provide clear semantics on resource access and sharing.

Monitors are small software components that conveniently \emph{retrofit}~\cite{overshadow, cloudvisor} new isolation boundaries into the existing software stack with minimal disruption by running at higher privilege levels, \eg through virtualization~\cite{overshadow, trustvisor, cloudvisor, hyperenclave}.
They strive to remain passive unless their isolation services are used, leaving existing software execution mostly unchanged, \eg they do not replace the kernel for process isolation.
They are popular for both confidential computing~\cite{cloudvisor, overshadow, komodo, veil, blackbox, hyperenclave} and encapsulation~\cite{erebor, secvisor, hypervision} to enforce fine grained access control.

Put together, a monitor that implements capabilities for access and boundary control can provide a single abstraction to compose isolation, support fine-grained heterogeneous trust models, and remain compatible with existing software and hardware stacks.
This is what we design and implement with \system{}.

Unlike systems in \autoref{tab:comparison-matrix}, \system{} covers all scenarios with a single enforcement layer, provides coherent attestation across multiple (potentially nested) \td{}s, and goes beyond the limitations of prior work~\cite{tdx-partitioning, veil} by supporting private shared memory between TEEs without imposing a cap on the number of isolated sub-components.
Because its semantics are decoupled from the underlying platform, \system{} achieves this uniformly across hardware; the single enforcement layer further reduces deployment complexity, making \system{} both general and portable.

\section{\system{} Overview}
\label{sec:overview}

\begin{figure*}[t]
    \includegraphics[width=\textwidth]{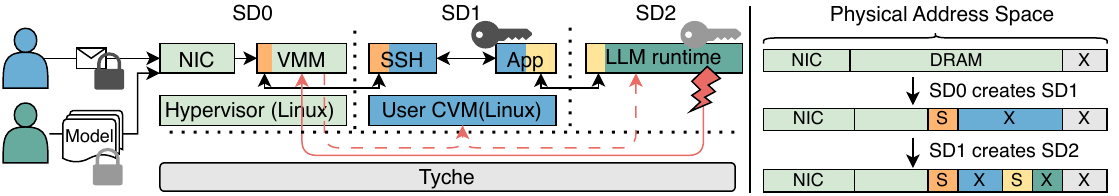}
    \caption{
      Mutual-distrust LLM inference deployment on top of \system{}.
      The left side shows the participating \td{}s -- \td{0} hypervisor, \td{1} CVM, and \td{2} nested enclave.
      The right side shows the machine’s physical memory address space as \td{}s are created;
      the colors indicate the regions accessible to each \td{} on the left, distinguishing exclusive regions (X) from shared ones (S).
      The encrypted model, prompts, and replies traverse memory along the black path, while interrupts are routed as shown by the red arrows.
    }
    \label{fig:llm-overview}
\end{figure*}

In this section we present an overview of \system{}'s design, its threat model, and an example deployment in \autoref{fig:llm-overview} that we will use throughout the paper.

\subsection{Architecture}

The \system{} security monitor exposes a capability-based API for creating and isolating security domains (\td{}s), its core abstraction.
All software other than the monitor runs inside an \td{}, an execution environment whose configuration and owned capabilities define its access to machine resources -- cores, devices, memory regions, and interrupts -- attested by \system{} as either shared or exclusive.
An \td{} can create child \td{}s, partitioning or sharing subsets of its resources and transferring control to children on selected cores through capability operations.
Rather than catering to a fixed trust model, \td{}s provide a unified, attestable abstraction for controlling shared and exclusive resources, accommodating a wide range of security policies, such as sandboxes, enclaves, and CVMs, and enabling their composition.

The API decouples resource management from the attestable enforcement of access restrictions.
Capabilities represent resources and govern \td{} interactions, letting domains implement their own isolation boundaries and scheduling policies.
\system{} tracks resource allocation and attests whether resources are shared or exclusive, capturing and enforcing dependencies between \td{}s.
It guarantees parents can reclaim resources or cores, while children are assured exclusive resources remain so unless explicitly shared, and that revocation or control transfer do not leak information.

Local and remote attestation make \td{} relationships explicit and verify end-to-end guarantees.
Attestation has two parts:
(1) a TPM-rooted measurement of the boot process and a public key, binding an \system{} binary to a physical machine and ensuring it runs alone at the highest privilege; and
(2) a report generated by \system{} and signed with the private key describing the \td{}’s resources, how they are shared, delegated to children, or can be reclaimed.

\system{}'s support for composable isolation enables controlled sharing.
\td{}s unify compartmentalization by supporting both confidentiality and encapsulation.
Confidentiality and integrity derive from exclusive access; encapsulation is enforced by restricting a child \td{}'s interactions and overlap of resources to trusted peers.
Recursive \td{} construction enables composition, letting each domain define its own boundaries.

\system{}'s platform-independent capabilities make it portable across architectures, with hardware-specific backends enforcing their semantics efficiently.
On Intel, \system{} uses \vtx{}~\cite{vtx}'s extended page tables and the \iommu{}~\cite{vtd}; on \riscv{}, it runs as \mmode{} firmware using Physical Memory Protection~\cite{pmp}.
The \framework{} adds a kernel driver to Linux environments to build nested sandboxes, enclaves, and CVMs using \td{}s.
This compatibility with existing hardware and software stacks ensures \system{}'s practicality.

\subsection{Threat Model}

\label{sec:threat-model}
\system{} assumes the underlying hardware, including physical devices and access control mechanisms, is trusted and part of its TCB.
The CSP and tenants are adversarial. 
We consider an attacker, running arbitrary code within an \td{}, restricted to an authorized subset of the machine's resources, such as cores, memory, and device configuration space.
In particular, the attacker might try to exploit: (1) the monitor's API, (2) device configuration space, and (3) privileged instructions to, e.g., emit or disable interrupts.
The attacker aims to access resources or \td{} state outside of its or the device's authorized sets, compromise the monitor's metadata, steal its private key, or hog resources to prevent revocation.

\myparagraph{Out-of-scope}
\textbf{Physical attacks}, such as accessing DRAM or the PCI bus to read the monitor or an \td{}'s memory, are out of scope of the current implementation, although they could be mitigated with hardware support, \eg{} total memory encryption~\cite{mktme} and PCI bus encryption~\cite{tdisp}.
\textbf{Side-channel}-based attacks are not explicitly addressed by \system{}, beyond appropriate flushes upon transitions, as it does not track shared micro-architectural state.
They however can be mitigated within the current implementation through core partitioning~\cite{coreslicing, coregapping} and physical memory allocation based on cache-coloring~\cite{scvirtualghost, marghera}.
\textbf{Denial-of-service} attacks to exhaust the monitor's memory are possible, but they only prevent the creation of new \td{}s and do not prevent the revocation, attestation, or isolation enforcement of existing ones.
On \amd{}, they can be mitigated by requiring \td{}s to supply memory for the monitor's metadata.
Inherent to the cloud, the CSP can deny service by turning off the platform~\cite{haven}.

\subsection{Running Example}

\label{sec:example}

\autoref{fig:llm-overview} shows a public cloud deployment where a user performs LLM inference with a proprietary model.
The threat model involves full mutual distrust: the user does not trust the CSP or model owner with prompts, the model owner does not trust the CSP or user with model weights, and the CSP does not trust either party.
Private inference in this setting requires confidential computing to protect prompts and weights, composed with encapsulation to prevent the LLM runtime from leaking prompts.

At boot, \system{} controls all resources and assigns them via capabilities to \td{0}, the CSP’s Linux+KVM~\cite{kvm} hypervisor running \framework{}.
To instantiate the user CVM, \td{0} uses the SDK to partition its memory, creating \td{1} with exclusive memory (blue in \autoref{fig:llm-overview}), shared memory (orange), and CPU cores.
\td{0} then transfers control to \td{1} on those cores.
Once booted, \td{1} (Linux + \framework{}) creates \td{2} for the LLM runtime in userspace, partitions its exclusive memory, and transfers a region (dark green) to \td{2}, forming an enclave isolated from both the CSP and the CVM but encapsulated by \td{1}.
The model owner attests \td{2}’s exclusive memory before provisioning the encrypted model.
Communication uses shared memory: \td{0}--\td{1} share a region for VIRTIO~\cite{virtio} networking (orange); \td{1}--\td{2} share a private region (yellow), not accessible by \td{0}, to exchange plaintext prompts and replies. Encrypted prompts arrive at \td{1} via orange, are decrypted into private memory (blue), and forwarded to \td{2} via yellow; replies follow the reverse path, encrypted by \td{1} before being sent to the VIRTIO network.
As there is no direct communication between \td{2} and \td{0}, \td{1} fully encapsulates the untrusted LLM runtime in \td{2} and thus prevents prompt leakage.
The model is delivered encrypted via orange, passed to \td{2} through yellow, and decrypted in its private memory (dark green).

All parties use \system{}’s remote attestation:
(1) The CSP confirms it retains resource control and manages interrupts (red arrows).
(2) The user confirms \td{1} is isolated from \td{0} and encapsulates \td{2}, which only communicates with the CVM.
(3) The model owner confirms \td{2} runs in exclusive memory without leaking the model.
(4) The user and model owner confirm interrupts do not leak data and that \system{} zeroes memory before returning it to the CSP.

\section{\system{} Design: API \& Capabilities}

\label{sec:capabilities}

\begin{table}[b]
{
    \small
    \renewcommand{\arraystretch}{1.0}
\begin{tabular}{|l|l|}
\hline
  Call & Description                                 \\ \hline
\texttt{CREATE}                   & Create a security domain                        \\ \hline
\texttt{SET/GET}                  & Set/Get a security domain's register or policy  \\ \hline
\texttt{SEND}                     & Transfer a capability ownership to an \td{}      \\ \hline
\texttt{SEAL}                     & Seal a security domain                          \\ \hline
\texttt{ATTEST}                   & Attest a security domain                        \\ \hline
\texttt{ENUMERATE}                & Discover info. about owned capabilities      \\ \hline
\texttt{SWITCH}                   & Control transfer into a security domain      \\ \hline
\texttt{ALIAS}                    & Create new memory region by aliasing one     \\ \hline
\texttt{CARVE}                    & Create new memory region by carving one      \\ \hline
\texttt{REVOKE}                   & Revoke a capability's child                  \\ \hline
\texttt{GETCHAN}                  & Create a channel from an \td{} or existing channel \\ \hline
\end{tabular}
}
    \caption{\system{}'s API}
\label{tab:monitor-api-simplified}
\end{table}

This section addresses our first research question by designing a small set of primitives to compose isolation and support heterogeneous trust models.
We focus on primitives for access and boundary control with clear, predictable semantics.
To achieve this, we devise software capabilities centered on sharing, transferring, and reclaiming resources.
Our capabilities make shared access and interactions explicit while decoupling policies from mechanisms~\cite{lampson} for portability across hardware platforms.
\system{} capabilities encode enough information to attest how an \td{} was created, what interactions it may have with others, and under which conditions it can be revoked.
This explicitness and fixed semantics gives the system predictable behavior, enabling long-term reasoning about isolation guarantees and trust relationships from a single attestation.

\system{} capabilities are unforgeable tokens issued by the monitor.
They are owned by security domains (\td{}s) and mediate access to two object types: memory regions and \td{}s.
Memory capabilities grant access to physical memory ranges, while \td{} capabilities govern \td{} management, \ie creation, configuration, attestation, and execution.
CPU cores and interrupts fall under \td{} capabilities, and devices are modeled as \td{}s.
In this section, we show that these two capability types and the narrow set of operations in \autoref{tab:monitor-api-simplified} to derive new capabilities are enough to isolate \td{}s and manage their nuanced trust relationships.

\subsection{Capability Derivation Trees}

\system{} maintains two capability derivation trees (CDTs): one for memory regions and one for security domains.
CDTs, widely used in capability systems~\cite{eros, sel4}, derive each capability from an existing one, inheriting equal or reduced access.
Derived capabilities appear as children of their parent node in their trees.
CDTs ensure (1) monotonicity of operations, (2) a record of operations, and (3) efficient cascading revocation of an entire subtree by revoking its root.

In \system{}, we further leverage the CDT structure to track overlapping resource access and dependencies between \td{}s, ensuring these relationships are reflected in attestation and that operations always leave the system in a consistent state.
The region CDT tracks exclusive and shared access to physical memory, while the \td{} CDT encodes \td{} management hierarchies, scheduling, and interrupt routing.
Both trees support efficient revocation and explicitly encode access policies and inter-object dependencies.

\subsection{Memory Region Capabilities}
\label{sec:memory-region-capabilities}

\system{} region capabilities offer a compact, attestable representation of shared or exclusive access to physical memory ranges, with security \emph{attributes} that define guarantees upon revocation.
\system{} initializes the CDT with a root memory region marked as \textbf{exclusive}, defined by a start and end address, access rights (read, write, execute), and an empty set of attributes.
\system{} transfers this root region's ownership to \td{0}, the first domain to run on the machine.

\begin{figure}
    \includegraphics[width=\columnwidth]{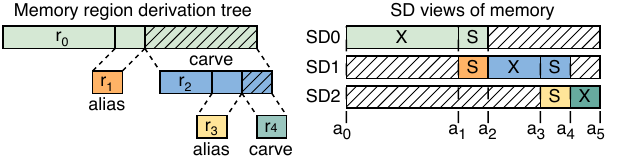}
    \caption{Memory region derivation tree and \td{} memory views based on \autoref{fig:llm-overview}. New regions are created by carving and aliasing an existing region. Sends between \td{}s are omitted. (S)hared and e(X)clusive memory is reported on views.}
    \label{fig:region-tree}
    \vspace*{-8pt}
\end{figure}

\myparagraph{Memory operations}
An \td{} creates new region capabilities from owned ones using \system{}'s \emph{alias} and \emph{carve} API calls (\autoref{tab:monitor-api-simplified}).
Alias creates a child region capability, marked as \textbf{aliased}, for a subrange of the parent region's physical addresses with equal or reduced access rights while preserving the parent's access.
A carve similarly creates a subregion but removes access to it from the parent region (see \autoref{fig:region-tree}, hatched area).
If the parent region was \textbf{exclusive}, the carved region is exclusive; otherwise, it is marked as aliased.
Carving enables confidential memory: an exclusive region stems from an unbroken chain of carves and no operation outside its subtree can alter its exclusivity.
\system{} ensures a capability's access rights and exclusivity are determined locally, based on its initial range, exclusivity status, and direct children.
In \autoref{fig:region-tree}, $r_0$ initially grants exclusive access from $a_0$ to $a_5$.
After an alias ($r_1$) and a carve ($r_2$), it retains exclusive access only from $a_0$ to $a_1$ and shared (aliased) access from $a_1$ to $a_2$, regardless of further subdivisions of $r_1$ or $r_2$.

\myparagraph{Memory management}
\system{}’s \emph{send} and \emph{revoke} calls let \td{}s transfer and reclaim memory.
\emph{Send} transfers ownership of a region, allocating it to the receiver.
\emph{Revoke} acts on a parent region to undo a child alias or carve, letting the parent’s owner reclaim memory sent via that child.
Revocations cascade, deleting the entire subtree rooted in the revoked capability: revoking $r_2$ from $r_0$ also revokes $r_3$ and $r_4$, restoring $r_0$’s access from $a_2$ to $a_5$.
To provide security guarantees to the receiver, \system{} allows the sender to optionally attach \emph{attributes} to the transferred region:
(1) \emph{hash} captures a cryptographic hash of an exclusive region's content, enabling the receiver to ensure it contains the correct initial data;
(2) \emph{clean} ensures the region is zeroed upon revocation to prevent data leakage; and
(3) \emph{vital} revokes the receiver if the capability is revoked, enforcing a minimal memory set necessary for functionality.
Attributes are set only when sending to an unsealed receiver (\autoref{sec:td-capa}), are bound to ownership rather than the CDT, and are non-monotonic.

The receiver then inspects received regions using \emph{enumerate} or \emph{attest}.
These calls report a region’s status (exclusive or aliased), initial range, access rights, attributes, and direct children, whether owned or not, as shown for $r_2$ in \autoref{fig:attestation}.

\begin{figure}
\small
    \begin{attestation}
sd1 = domain {r1, r2, sd2}
      |registers.HASH: 8988ef57...
      |cores: 0b11
      |mon.api: 0b11111111111 | RECEIVE
      |interrupts: {
      | 0 -> {Report, registers: 0b0},
      |  ...       }
sd2 = domain {r3, r4} 
      |registers.HASH: 978de00f...
      |cores: 0b01
      |mon.api: 0b00001110000 | !RECEIVE
      |interrupts: {
      | 0 -> {Not report, registers: 0b0},
      | ...        }
r1 = aliased a1 a2 with RW_
r2 = exclusive a2 a5
     with RWX, HASH|CLEAN|VITAL
     |HASH: 755ee2b2...
     |alias at a3 a4 for r3 with RW_
     |carve at a4 a5 for r4 with RWX
signature: a0e0d23f26564bd5...
    \end{attestation}
    \caption{ 
      Simplified attestation for \td{1} with \autoref{fig:region-tree}'s nomenclature.
      Allowed monitor API calls are encoded as bitmaps based on the order from \autoref{tab:monitor-api-simplified} (\eg bit 0 is create). 
    }
    \label{fig:attestation}
    \vspace*{-6pt}
\end{figure}

\subsection{Security Domain Capabilities}
\label{sec:td-capa}

\system{} initializes \td{0}, the first \td{} which owns the root memory region, runs on all cores, and handles all interrupts; all other \td{}s are created from subsets of these initial resources.

\myparagraph{\td{} creation and configuration}
\emph{Create} enables an \td{} to spawn a child in the CDT and obtain an \td{} capability referencing the newly created domain.
The capability enables the parent to configure, send regions, seal, attest, schedule, and revoke the child.
The child is initially \emph{unsealed} and cannot execute.
The parent configures the child via \emph{set}, specifying per-core registers state and \td{} \emph{policies}.
Policies define the child's allowed cores, permitted monitor calls and whether they are allowed from user space, whether it can receive new capabilities after \emph{sealing}, and interrupt policies.
Policies are monotonic: \system{} rejects \emph{set}s exceeding the parent’s rights.
Memory is provisioned as in \autoref{fig:creation-td1}, with the parent aliasing and carving its regions before sending them.
\emph{Seal} makes the \td{} executable and prevents further sets to its registers and policies.
Revoking an \td{} triggers the cascading revocation of all its owned capabilities and its CDT subtree.

\myparagraph{\td{} execution \& interrupts}
Control transfers between \td{}s \ie transitions on a core, occur via explicit monitor calls or upon interrupts.
The \emph{switch} API call implements a call-return model: \system{} ensures the callee is authorized to run on the core, saves the caller's state, loads the callee, and transfers control.
Switching to a child requires its \td{} capability; a switch with no \td{} argument returns to the parent.

Interrupts trigger \td{} control transfers. They propagate through the CDT, which encodes routing and handling policies.
For each interrupt vector, \td{}'s interrupt policies specify whether to \emph{Deliver}, \emph{Report}, or \emph{Not report} the interrupt, and which registers the parent can access during handling.
An interrupt marked as \emph{Deliver} is received by the \td{} directly.
When an interrupt not marked as \emph{Deliver} occurs, \system{} preempts the \td{} and walks the CDT upwards to transfer control to the first \td{} (the handler) it finds with a \emph{Deliver} policy.
To the handler, the interrupt appears as a return from a switch to its direct child, with the interrupt acting as the return value.
After taking care of the interrupt, the handler resumes the child's execution by performing a switch.
The return path walks the CDT downwards, with \system{} reporting the interrupt to all \td{}s with a \emph{Report} policy in a manner similar to the original handler \td{}.
These \td{}s observe the interrupt as if it originated from their direct child and decide whether to resume execution.
Those with a \emph{Not report} policy are skipped.

Consider the deployment in \autoref{fig:llm-overview}.
\td{2} triggers a divide-by-zero exception (interrupt vector 0).
Per \autoref{fig:attestation}, \td{2}'s policy for vector 0 is \textit{Not report}, so the exception traps to \system{}, which walks up the CDT searching for an ancestor \td{} with a \textit{Deliver} policy.
Skipping \td{1}, whose policy is \textit{Report}, \system{} switches to \td{0}---the first ancestor with a \textit{Deliver} policy---presenting the event to \td{0} as a return from a switch to \td{1}.
\td{0} handles the exception as if originating from \td{1}'s subtree, with access to \td{1}'s registers restricted by \td{1}'s policy, then switches back to \td{1}.
Because \td{1} is marked \textit{Report}, \system{} delivers the return to \td{1}, which sees it as a return from a switch to \td{2} caused by a divide-by-zero.
Before resuming \td{2}, \td{1} may inspect or modify \td{2}'s register state---within the bounds defined by its policy---and decide how to proceed, for example resuming \td{2} after correcting the faulting state.
This routing and return protocol enables flexible interrupt delivery for isolation abstractions ranging from sandboxes to nested VMs, while providing attestable scheduling guarantees on how \td{}s can be preempted.

\myparagraph{\td{}s interactions}
\td{}s communicate via shared memory, control transfers to direct children, and interrupts.
To ensure that parents remain responsible for scheduling and revoking their children, \td{} capabilities cannot be transferred between \td{}s.
To enable non parent-child \td{}s to attest each other and exchange regions without relying on a common ancestor, \system{} supports \emph{channel} capabilities.
Channels are derived from \td{} capabilities using \emph{getchan} and appear as children in the CDT.
They act as weak references to \td{}s and can be transferred between \td{}s via \emph{send}.
Channels allow non parent-child \td{}s to directly attest each other, exchange memory regions (\eg to establish private shared memory), or share other channels -- but not to schedule, configure, or revoke an \td{}.

\myparagraph{Devices}
\system{} models devices as \td{}s, with configuration space, DMA, and port I/O access mediated through region capabilities.
Interrupt routing follows the same protocol as above, but \system{} allows backends to optimize core-routing using platform specific hardware mechanisms.
\td{}s on the CPU interact with devices through shared regions for the device's configuration space and MMIO.
\system{} provides \td{0} with a channel to every device.
\td{0} delegates device access to an \td{} by carving the configuration space and duplicating a channel, sending both to the \td{} to enable direct \td{} to device interactions.

\begin{figure}
    \includegraphics[width=\columnwidth]{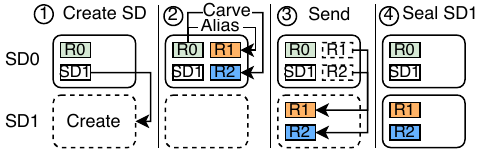}
    \caption{Capability operations to create and configure \td{1}.}
    \label{fig:creation-td1} 
    \vspace*{-4pt}
\end{figure}

\begin{figure}
    \includegraphics[width=1\columnwidth]{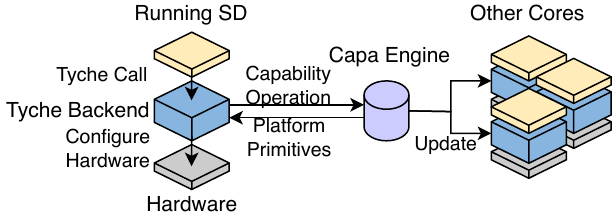}
    \caption{The capability engine maintains the system state across all cores.}
    \label{fig:capa-engine}
    \vspace*{-4pt}
\end{figure}

\subsection{Attestation \& Security guarantees}

\td{}s can request an attestation for themselves and any \td{} for which they own an \td{} capability (or channel).
\system{}'s attestation reports the \td{}'s configuration, including a hash of initial register content, policies, and a description of owned capabilities.
For remote attestation, an \td{}'s \emph{attest} call includes the remote verifier's public key and nonce, as well as a public key generated by the domain itself, ensuring they are all measured and signed by \system{}.

Conceptually, the attestation provides, for each owned capability. visibility into its direct children in their respective CDT, but not further.
\autoref{fig:attestation} shows the attestation of \td{1}.
It reports the configuration of \td{1} and its child \td{2}, along with their owned capabilities.
Allowed cores, monitor calls, and registers that can be read and written upon interrupts are encoded as bitmaps.
Lineage between regions is made explicit through naming: \eg $r_3$, owned by \td{2}, is derived from $r_2$.
If \td{2} held a capability not derived from \td{1}, it would appear only in \td{2}’s set of owned capabilities with a fresh name and no further information.
Indices are reused here for clarity; in practice, \system{} assigns fresh names, starting from 0 based on the requester’s capabilities, to avoid revealing information about the broader system.

\system{}'s attestation makes sharing and interactions explicit, while providing guarantees on initial \td{} state, run time events, and resource reclamation.
\system{} thus supports confidentiality and integrity, encapsulation, or even the privacy of communications between distrustful \td{}s.
Confidentiality and integrity are ensured by exclusive ownership of regions and the absence of leakage through interrupts or revocation (\ie attributes).
Encapsulation, as for \td{2} in \autoref{fig:attestation}, is achieved by ensuring a child's regions are a subset of the parent's exclusive ones, and that its policies prevent receiving or sending capabilities after sealing.
The child can still be confidential, as is the case for \td{2}, and \td{1} can safely share information with it via their private shared memory in $r_3$ without risk of leakage or trusting \td{2}'s implementation.

\section{\system{} Implementation}

\label{sec:implementation}

\autoref{fig:capa-engine} shows the \system{} security monitor has two components:
(1) a platform-independent \emph{capability engine} and
(2) a platform-specific \emph{backend}.
The backend translates hardware events, such as interrupts or calls to \system{}, into capability operations and forwards them to the capability engine, shared across cores.
The capability engine implements the system's global capability state machine, validates and executes operations and notifies the backend of configuration changes on affected cores via \emph{updates}.
The capability engine and backend are part of the TCB.

To construct a backend and provide the attestable enforcement of isolation by the monitor and its capability engine, the hardware must provide:
(1) the ability to establish trust in the monitor;
(2) the monitor's exclusive oversight of an access control mechanism to enforce resource isolation (memory, CPU, interrupts, devices);
(3) a direct, secured communication channel between \td{}s and the monitor.

This section describes the custom loader we use on bare metal to establish trust in the monitor, the capability engine, details the access control enforcement and communication in the \amd{} backend, and provides an overview for \riscv{}.
\subsection{\system{} attested boot}

\system{} boots via a custom bootloader that enumerates CPU cores, devices, and DRAM to generate a \emph{\manifest{}}.
It reserves memory for \system{} per a compile-time configuration, loads the monitor (capability engine and backend), and maps the \manifest{} into its address space.
The bootloader generates an attestation key pair, measures the monitor, \manifest{}, and public key into a TPM~\cite{tpm} PCR, and transfers control to the monitor with the private key.
Using the \manifest{}, the backend initializes the capability engine, creates \td{0} (\eg a stock Linux kernel) with access to all unused memory and device regions, and sets up an I/O \td{} for each DMA-capable device with an alias to \td{0}’s memory.
Control is then transferred on all cores to \td{0}.

Like prior work~\cite{trustvisor}, an \td{} attestation includes a TPM quote and a domain attestation signed by \system{} (\autoref{sec:capabilities}).
The quote binds the monitor binary, platform, and attestation key via PCR values, enabling trust in the monitor and transitively in \td{} attestations.
The attestation covers the full boot chain, including our bootloader.
We also prototyped dynamic root of trust (DRTM) support~\cite{txt, amd-skinit} via Intel TxT~\cite{txt}, reducing the TCB to just the capability engine and backend.

\subsection{Capability engine}
\label{sec:capa-engine}

The capability engine implements the region and \td{} capability derivation trees (CDTs from \autoref{sec:capabilities}), validates and executes capability operations, and computes \emph{updates} supplied to the backend to reflect configuration changes onto the hardware.
It is implemented as a standalone, bare-metal (\texttt{no-std}) Rust library.
The engine consists of \nbloccapa{} lines of code, uses no \texttt{unsafe}~\cite{safe-rust} to ensure memory safety, and is fuzzed as part of our continuous integration (CI) setup using LLVM's \emph{libFuzzer}~\cite{libfuzzer} to proactively detect and fix bugs.

The capability engine exposes the API defined in \autoref{sec:capabilities} for capability operations and integrates within a backend via a \emph{platform} interface (Rust trait).
This interface is supplied by the backend to provide platform-specific primitives to manage per-\td{} platform state (\eg per-core registers and access control mechanism configuration such as extended page tables (EPTs)), map and unmap physical memory ranges in an \td{}'s platform state, manage interrupts, and implement cross-core synchronization primitives in the form of IPIs, barriers, and locks. 

At all times, the engine tracks which \td{} executes on each core and ensures a consistent state view across cores.
The engine uses the CDTs to determine which \td{}s and regions an operation affects and serializes operations with overlapping targets to ensure consistency.
For each operation, the engine derives the set of affected cores and enqueues an update reflecting the result of the operation in per-affected-core queues, as shown on \autoref{fig:capa-engine}.
The engine uses the platform interface to preempt affected cores (IPIs), requesting them to process the update and block on a barrier while the initiating core makes the new state globally visible.
They are then unblocked and apply the changes to their local hardware state.
This ensures the atomicity of capability and hardware state changes across cores.
In practice, only a small set of updates are needed: access right changes, \td{} revocation, and interrupt delivery. 

For example, sending a carve is handled by a core in the engine, which validates the request, computes the new capability state, and enqueues an access-right update on all cores running the sender or receiver.
Using the platform IPI, it preempts these cores so they can process the update.
The update involves two synchronization barriers: the first ensures all affected cores are preempted; the second waits for the new platform state.
Between them, the initiating core updates the platform state -- unmapping the region from the sender and mapping it into the receiver (\eg via EPTs on \amd{}).
Finally, it finalizes the capability state and releases the second barrier, letting other cores apply the new state locally (\eg TLB shootdown).

The capability engine requires memory to allocate capability metadata and \td{} state from.
\system{} provides this memory in two ways: statically or dynamically.
In the static configuration, \system{} reserves a contiguous block at boot and passes it to the engine.
On \riscv{}, this is required because the architecture provides only a small number of PMP entries, and \system{} must protect itself and its metadata using at most one region (\S\ref{sec:riscv}).
To prevent memory exhaustion from becoming a covert channel or denial-of-service vector, \system{} can enforce per-\td{} capability quotas.

On platforms without this restriction, \eg using EPTs on \amd{}, \system{} supports dynamic memory management in which \td{}s supply memory for their children’s metadata~\cite{tock, sel4}.
This is integrated into region capabilities semantics with a \emph{META} attribute (\S\ref{sec:memory-region-capabilities}).
Regions transferred as \emph{META} to a child must be exclusive, \emph{vital}, and are \emph{clean}ed (zeroed) on revocation.
The monitor uses these regions to host, \eg a child’s EPTs.
They are inaccessible to both parent and child, but are included as other regions in attestations.

Both approaches introduce trade-offs.
Static reservation exposes \system{} to denial-of-service via resource exhaustion, though isolation, attestation, switches, interrupts, and revocation always succeed (\S\ref{sec:threat-model}).
Dynamic management avoids this limitation but enables potential cache-based side-channel attacks on metadata, which should be mitigated with allocation based on cache coloring -- hence their inclusion in attestation.

\subsection{\system{} \amd{} backend in root mode}
\label{sec:amd-backend}

Intel VT-x~\cite{vtx} accelerates virtualization by running a host in \texttt{root} mode and guest VMs in \texttt{non-root} mode.
A virtual machine control structure (VMCS) governs guest execution, virtualizing privileged operations like \texttt{cr3} writes without host intervention.
Extended page tables (EPTs)~\cite{vtx, amd-npt} restrict non-root memory access.
Guests trap to the host via the \texttt{vmcall} instruction.

The \system{} \amd{} backend reuses VT-x to isolate \td{}s, running in root mode and executing \td{}s in non-root mode.
Each \td{} has its own EPT (enforcing region capability access) and per-core VMCS.
Devices are also modeled as \td{}s, with memory isolation enforced via Intel \iommu{}~\cite{vtd}.
\system{} uses \texttt{INIT IPI}s to trap cores into the monitor and implements synchronization with semaphores, atomics, and spinlocks.
The backend is 6K lines of Rust, bloated by error codes and VMCS field enums, with unsafe code for hardware setup.
Along with the capability engine, it compiles to only \nbruntimebytes{}. 

\td{}s interact with the monitor via the non-interposable \emph{vmcall} instruction.
Arguments and return values are passed through general-purpose registers, with capabilities referenced by \td{}-local indices (like UNIX file descriptors).
Large values (\eg attestations) use a multi-\emph{vmcall} protocol.
We measured register-based communication to be efficient, supporting tens of thousands of large (4KB) attestations per second.
Registers avoid monitor access to \td{} memory, reducing risks like confused deputy~\cite{confused-deputy} and cache attacks.
The monitor accesses memory only when no \td{} can, to enforce \texttt{Clean} (zeroing) or \texttt{Hash} (reading).
Future work aims to offload even these to more restricted environments~\cite{sanctum}, ensuring complete isolation.

\td{} switches save the caller’s VMCS and load the callee’s.
The backend manages general-purpose registers in software and handles other state, including MSRs, using a mix of hardware and software to prevent leakage.
EPT TLB entries are tagged with virtual processor IDs to reduce flushes on transitions.
An \td{} runs until it returns or receives an unhandled interrupt.
All control transfers---whether through \texttt{switch} or interrupts---go through the monitor, which tracks per-core state and enforces \td{} policies.

\td{} interrupt policies are offloaded to hardware via VMCS where possible.
Intel \vtx{} allows fine-grained traps for exceptions (vectors 0--31) but only a single bit for external interrupts.
Thus, external interrupts trap to the backend which selectively decides to either re-inject them or switch to an ancestor \td{} per \autoref{sec:capabilities}.
When supported, the \amd{} backend uses hardware APIC virtualization and \iommu{} interrupt remapping to reduce traps and gain finer control.
It can also expose VMCS fields as \td{} registers, letting parents manage interrupt delivery at the cost of limiting child policies.

\subsection{\system{} \riscv{} backend in \mmode{}}
\label{sec:riscv}

On \riscv{}, \system{} runs in machine mode (\mmode{}), and the backend uses Physical Memory Protection (PMP)~\cite{pmp} to enforce region capabilities.
PMP entries enforce permission on contiguous segments of physical addresses and can only be configured from \mmode{}.
Each core has a limited number of PMP registers (up to 64), with one reserved to protect \system{} itself.
If an \td{} configuration cannot be satisfied with the available PMP registers, a synthetic exception is injected and delegated to a parent \td{} following the interrupt routing protocol.
As capability revocations always succeed, the parent can recover regions sent to the child.
Like \amd{}'s \iommu{}, \system{} requires IO-PMPs~\cite{iopmp, siopmp} to defend against rogue DMA requests.
Like \amd{}, interrupts and exceptions can be either trapped or delivered directly to an \td{} by configuring the \texttt{mideleg} and \texttt{medeleg} registers.
Communication with the monitor occurs via \emph{ecalls} (for S-mode) or illegal instruction traps (for U-mode), as S-mode manages \emph{ecalls} from U-mode within an \td{} to provide fast system calls.

Running in \mmode{} imposes more constraints than \amd{} virtualization-based backend,
as PMPs are only available in limited supply (8 on our board~\cite{starfive}).
The \td{}s must carefully manage memory to maximize the use of contiguous memory, like existing \riscv{} security monitors~\cite{keystone, cure, sanctum, coreslicing}.
Alternatively, future support for \hmode{} could enable a virtualization-based backend.

\section{Integrating Existing Software with \system{}}

\label{sec:deploy}

The \framework{} is a set of drivers and ports of popular software frameworks that let stock Linux \td{}s run unmodified workloads as sandboxes, enclaves, or CVMs in separate \td{}s.
These domains can be composed to secure complex cloud deployments.
The \framework{} is \textbf{not} part of \system{}'s TCB.

\subsection{Interfacing with \system{} from Linux environments}

We provide \textbf{\capadriver{}}, a kernel driver on \amd{} and \riscv{} that lets Linux environments in an \td{} interact with the monitor and create new \td{}s.
\capadriver{} abstracts capabilities and exposes a simple interface --\emph{ioctl} commands, file descriptors, and \texttt{mmap} -- to create, manage, and run \td{}s.
It allocates memory from kernel pools, reserves pages (as in ballooning~\cite{ballooning}), and ensures revocation to avoid leaks after crashes.
To reduce fragmentation and work around \riscv{}’s limited PMP entries, it can reserve contiguous physical memory at boot via kernel parameters.
\capadriver{} schedules \td{}s as part of their host process, like VMs under KVM~\cite{kvm}, and allows them to coexist with OS isolation tools -- cgroups, namespaces, and syscall filtering -- at the process boundary.

\subsection{Backward compatibility}
\label{sec:frameworks}

The \framework{} provides backward-compatibility with popular virtual machine monitors (VMMs) and enclave frameworks.
It relies on the \capadriver{} driver as a compatibility layer to run VMs, CVMs, sandboxes, and enclaves as \td{}s.

\myparagraph{VMs \& CVMs with KVM on Intel \amd{}}
\textbf{\kvmdriver{}} is a fork of the \kvmintel{}~\cite{kvm} driver, modified to run VMs and CVMs as child \td{}s (\textit{not} as nested VMs).
The porting effort involved modifying just 400 LOC out of 14.6k LOC, replacing Intel \vtx{} instructions and EPT management with \capadriver{} calls.
With this patch, \system{} can support popular KVM-based VMMs and container frameworks~\cite{qemu,lkvm,gvisor,firecracker}.

\kvmdriver{} maintains the runtime behavior of VMs compared to \kvmintel{} and thus uses \system{}'s ability to delegate APIC virtualization to a parent \td{}.
For CVMs, additional logic was added to account for the VM's memory and state being unavailable to the host.
\kvmdriver{} is backward-compatible with existing VMMs for non-confidential VMs.
A small 20-line patch to the LKVM~\cite{lkvm} VMM enables confidential VM support by requesting exclusive memory from \capadriver{}.

\myparagraph{Sandboxes \& enclaves on x86 with Gramine}

Gramine~\cite{gramine}, formerly Graphene~\cite{graphene}, is a library OS (libOS) for running applications inside Intel SGX~\cite{sgx} enclaves; it shields applications by mediating system calls to the untrusted OS.
In most cases, Gramine supports unmodified binaries.
\textbf{\graminetyche{}} is a fork of Gramine SGX platform abstraction layer that runs enclaves and sandboxes as \td{}s in userspace, isolating programs in exclusive or shared memory without Intel SGX hardware.
We modified 300 lines of code (LOC) out of 14.9k LOC, replacing Intel SGX logic with \capadriver{} calls.
	The SDK also populates the \td{}'s page tables: it allocates the corresponding physical pages, carves and sends them to the child \td{}, and initializes its page table root register (\texttt{cr3}).
	Once the enclave \td{} is sealed, its policy prevents the parent from reading or writing \texttt{cr3}, ensuring the parent cannot tamper with the child's address translation.

	\graminetyche{} does not yet support \texttt{fork}.
	This is a limitation of our port, as
	doing so requires non-trivial cross-enclave synchronization and state transfer, not an intrinsic constraint of \system{}.
For sandboxing, \td{}s complement OS-based mechanisms by providing efficient, hardware-enforced, and OS-independent isolation with a reduced TCB.
They enable attestation not only of the sandbox boundaries but also of the trusted code implementing higher-level policies (\eg syscall filtering).
Future SDK support could enable sandboxing within enclaves, which is beyond the capabilities of standard OS isolation.

\myparagraph{Enclaves on \riscv{} with Keystone}
Keystone~\cite{keystone} is a \riscv{} TEE framework, consisting of a Linux kernel driver and the Eyrie enclave runtime.
Porting Keystone to \system{} required 20 LOC changes to the runtime for compatibility with \system{}'s API, and 150 LOC to the Keystone driver to interact with \capadriver{} for creating and managing \td{}s.

\subsection{Preserving runtime behavior}

\framework{} preserves existing abstraction semantics, allowing fair performance comparisons (\autoref{sec:evaluation}).
Future work could better exploit \system{}'s features -- for example, running Gramine’s libOS as privileged software to reduce enclave exits~\cite{gramine-tdx, tdx}, or using \td{}s in a \system{}-aware VMM for efficient I/O and delegated interrupt handling.
Timer interrupts, currently routed to the hypervisor, could instead follow \td{} policies to avoid VM exits.

\section{Evaluation}

\label{sec:evaluation}

\begin{figure}
    \includegraphics[width=\columnwidth]{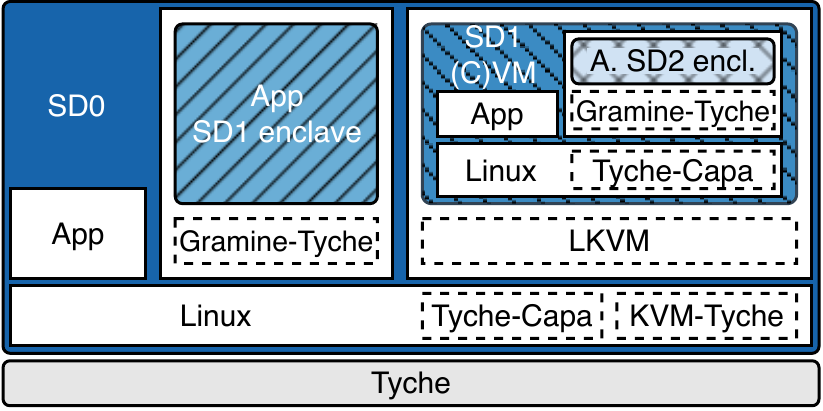}
    \caption{
      Superposed system and \td{} views of the deployments benchmarked in \autoref{fig:perf_comparison}.
      \td{}s are blue rounded boxes. System abstractions (processes, kernels, and VMs) are full rectangles, and libraries and drivers dotted ones. 
    }
    \label{fig:eval-config}
\end{figure}

This section evaluates \system{} design \& performance on \amd{}.
It reports microbenchmarks for the full-fledged \amd{} implementation and compares it with the prototype on \riscv{} (\autoref{sec:eval:micro}).
It then focuses on \amd{} to measure the performance of real-world applications isolated with various threat models (\autoref{sec:eval:programs}), using native execution as a baseline and including Intel SGX enclaves and native VMs as reference points.
Case studies (\autoref{sec:eval:use-case}) then detail the confidential LLaMa CPU-based inference for mutually distrustful model and prompt owners and other isolation deployments enabled by \system{}.

\myparagraph{Experimental setup}
On \amd{}, \system{} runs on a 16-core \cpuamd{} CPU with Intel SGX~\cite{sgx} v1 and an EPC of \epcsize{}, using a stock Linux kernel \linuxamd{} as \td{0}.
On \riscv{}, it runs on a \cpuriscv{}~\cite{starfive} (JH7110 SoC) with 8 PMP entries per hart and a stock \linuxriscv{} kernel.
VMs and CVMs use \linuxamd{} with 4 GiB, 8 cores, and VIRTIO devices~\cite{virtio}.
CVMs also enable SWIOTLB~\cite{swiotlb} to copy I/O into \td{0}-shared memory.
Network experiments use a separate 12-core client~\cite{memtier, wrk} with 400 connections over Ethernet.
Results average the best 9 of 10 runs.

\myparagraph{Naming conventions}
We refer to bare-metal Linux as "native" and KVM-based Linux VMs as "native VMs".
\td{0} runs Linux and creates \td{1} VMs, CVMs, and Gramine-based enclaves.
\td{2} enclaves are nested inside \td{1} CVMs and isolated from both \td{0} and their parent.
SGX enclaves run with Gramine-SGX.
Sandbox \td{}s perform identically to enclave \td{}s and are omitted. See \autoref{fig:eval-config}.

\subsection{Microbenchmarks}
\label{sec:eval:micro}
\label{sec:eval:overheads}

We measure the average latency (over 1000 ops) of \system{}'s primitives on both platforms to compare hardware mechanisms.
We then evaluate CPU and I/O overheads on \amd{} across \td{} configurations, compared to equivalent native deployment.

\begin{figure}
    \includegraphics[width=\columnwidth]{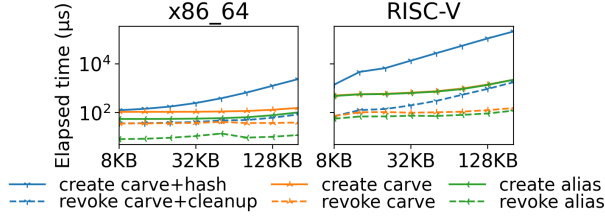}
    \caption{Create \& revoke latencies as a function of size for carve, with and without hash \& clean, and alias \td{} memory.}
    \label{fig:create-bench}
\end{figure}

\myparagraph{Creation/Revocation}
\autoref{fig:create-bench} shows carves are more expensive than aliases as they remove memory from \td{0} and notify other cores via IPIs (\autoref{sec:capa-engine}).
This difference is especially noticeable on \amd{}, where carves trigger a walk on \td{0}'s EPT.
On both platforms, \textit{hash} and \textit{clean} increase latencies with the \td{}'s size, as they require reading and writing memory, respectively.

\begin{table}[b]
    \center
\begin{tabular}{l|c|c|}
\cline{2-3}
                            & \textbf{Enter+Exit \system{}}  & \textbf{Total switch cost} \\ \hline
\multicolumn{1}{|l|}{\textbf{\amd{}}}   &    0.493 +/- 0.017 \microsecond & 1.171 +/- 0.002 \microsecond \\ \hline
\multicolumn{1}{|l|}{\textbf{\riscv{}}} &    0.246 +/- 0.000 \microsecond & 3.897 +/- 0.007 \microsecond   \\ \hline
\end{tabular}
\caption{Average switch latency and standard deviation, including monitor entry+exit (hardware privilege transitions).}
\label{fig:transition-cost}
\end{table}

\myparagraph{Switches}
In \autoref{fig:transition-cost}, \td{} switches on \riscv{} are 3x slower than on \amd{}, despite faster privilege layer transitions to/from \system{}.
On \amd{}, \system{} uses a hardware-software combination to efficiently save and restore \td{} states, while \riscv{} relies entirely on software.
Overall, \system{}'s switch latencies on \amd{} (\amdtrans{}) and \riscv{} (\riscvtrans{}) are competitive with related work and hardware extensions~\cite{sgx, tdx, graphene, veil, keystone}.

\begin{figure}
    \includegraphics[width=\columnwidth]{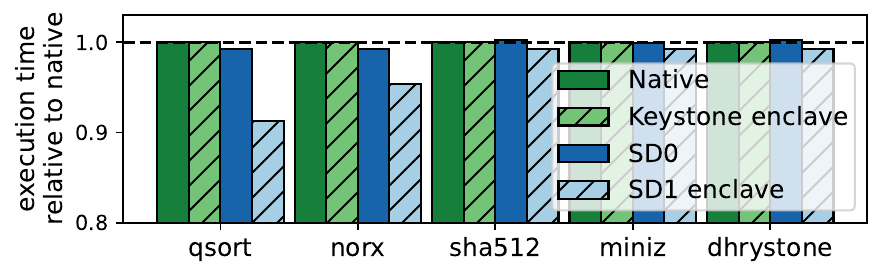}
    \caption{RV8 \riscv{} CPU microbenchmarks comparison between native, unmodified Keystone, and \system{}.}
    \label{fig:rv8}
\end{figure}

\myparagraph{\riscv{} prototype}
On \autoref{fig:rv8}, \system{} on \riscv{} competes with native and unmodified Keystone on most RV8 microbenchmarks, with a \rvslowdownqsort{} slowdown for \td{1} enclaves on short-lived programs (qsort, norx) due to the extra indirection to the \capadriver{} driver. 

\begin{figure}
    \includegraphics[width=\columnwidth]{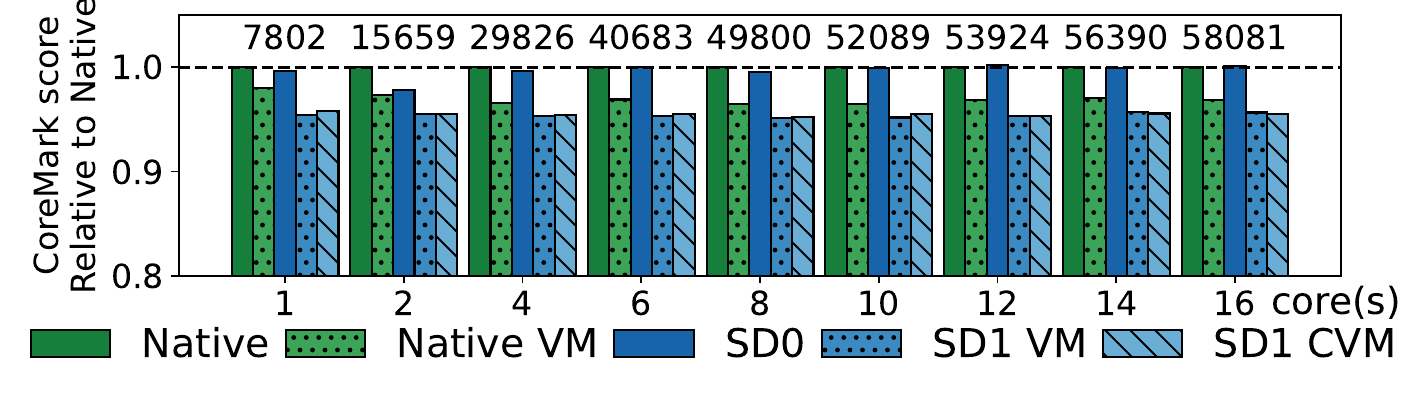}
    \caption{CoreMark-Pro performance relative to bare-metal Linux (Native) for different \td{} deployments and varying number of cores on \amd{} (raw native score at the top).}
    \label{fig:coremark}
\end{figure}

\myparagraph{CPU overheads on \amd{}}
\autoref{fig:coremark} shows CoreMark-PRO results from 1 to 16 cores.
\td{0} matches bare-metal, with minor slowdowns (up to \tdtwothreadslowdown{}) at 2-4-8 cores due to virtualization amplifying cache and TLB effects.
\td{1} VMs and CVMs show slight overhead: \tdvmcpuoverheadnativevm{} over native VMs and \tdvmcpuoverheadnative{} over native, mainly from timer interrupt transitions.
CVMs add no extra cost over VMs

\myparagraph{Multi-(C)VMs}
We group the 16 cores into 2-, 4-, or 8-core sets, launching one Linux (C)VM per group (\eg 8 VMs for 2-core groups).
Each VM boots a full Linux kernel, runs two-minutes \emph{sysbench}~\cite{sysbench} CPU and memory benchmarks, saves the results, and shuts down:

\lstset{
  basicstyle=\ttfamily\small,
  breaklines=true,
}

\begin{lstlisting}[
  basicstyle=\ttfamily\small,
  columns=fullflexible,
  keepspaces=true,
  breaklines=true
]
sysbench cpu --threads=$(nproc) --time=120 run
sysbench memory --time=120 --threads=$(nproc)
--memory-block-size=128M --memory-access-mode=rnd
--memory-oper=write run 
\end{lstlisting}

CPU throughput (events/sec) and memory throughput (random writes over 128 MiB) show only 2\% CPU and \textless{}1\% memory overhead for \td{}s versus native VMs.
\emph{kvm\_stat}~\cite{kvmstat} shows similar event distributions; the extra cost is due to monitor indirections on timer interrupts and synchronized access to \td{0} state.
Running two \td{} (C)VMs per group yields similar overheads compared to native VMs, showing they remain stable under oversubscription.

\begin{table*}[!t]
\small
\centering
\begin{tabular}{lcccccc}
\toprule
 Server   & Benchmark client & Threads & Connections & Workload configuration    & Pipeline & Termination condition \\
\midrule
 redis    & memtier          & 12      & 400         & 10\% write, 90\% read     & 1        & 20 million requests \\
 hyper    & wrk              & 12      & 400         & 12B payload + 76B headers & 1        & 2 minutes \\
 lighttpd & wrk              & 12      & 400         & 10kiB payload             & 1        & 2 minutes \\
\bottomrule
\end{tabular}
\caption{Network benchmarks configuration}
\label{tab:net-bench-config}
\end{table*}

\myparagraph{I/O overheads on \amd{}}
We measure Redis I/O using memtier~\cite{memtier} with pipeline=1 GETs to avoid masking latency and to expose \system{}’s overhead.
\autoref{tab:net-bench-config} details the full configuration of the networking benchmarks.
\autoref{fig:perf_comparison} and \autoref{fig:redis_latency} show throughput and latency across \td{} configurations.
Most native and \td{} (C)VM overhead comes from the LKVM user-space VIRTIO driver; running native and \td{} VMs with QEMU (no VHOST) improves throughput by 15\%-points, confirming this bottleneck.
With QEMU, \td{} VMs still achieve similar performance to native ones, even with VHOST enabled.
This is expected as better drivers (QEMU) and VHOST improve the performance on the common path for both native and \td{} (C)VMs and are orthogonal to \system{}.
Enclave overheads stem from libOS copies:
\td{2} in particular suffers from compounded I/O paths -- VIRTIO, SWIOTLB, and enclave copying.
The lack of pipelining degrades the throughput compared to the baseline due to the higher latency introduced by the different levels of isolation, as shown in \autoref{fig:redis_latency}.
In particular, the relative performance of native and \td{} VMs compared to native or \td{}0 in \autoref{fig:perf_comparison} is due to the increased latency: from 0.86ms when running natively to 1.77ms in a VM at p50 on the Redis benchmark.
 Enabling pipelining hides the latency, we measure a baseline 4.16 millions req/s with pipeline = 30, and 3.54 millions in the native and \td{} VMs, \ie{} a 15\% overhead rather than 40\%.
With lighttpd~\cite{lighttpd}, we observe that increasing payload size amortizes copy and context switching overhead, reaching native and \td{0} throughput for 10 KiB payloads.

\begin{figure}
    \includegraphics[width=\columnwidth]{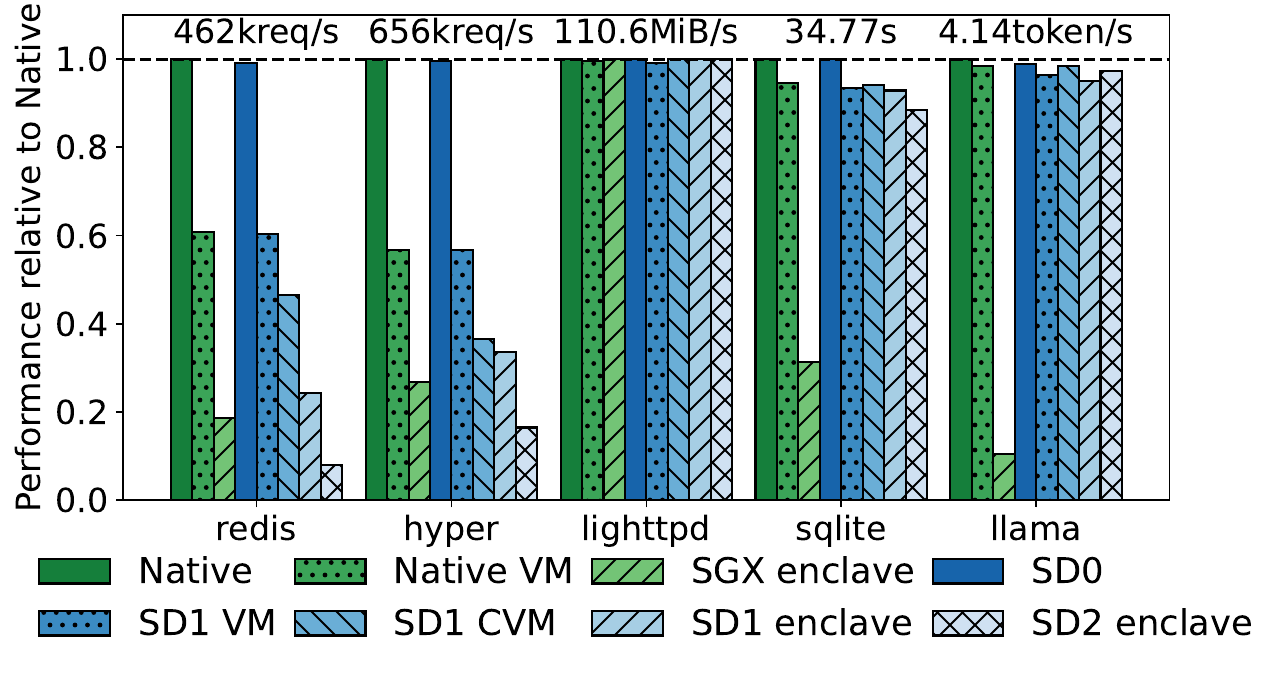}
    \caption{Performance relative to native execution (green) for real-world applications isolated as \td{}s on \amd{} (blue).}
    \label{fig:perf_comparison}
\end{figure}

\subsection{Enclaves, VMs, CVMs, \& composable isolation}
\label{sec:eval:programs}

\autoref{fig:perf_comparison} shows \system{}’s performance isolating real-world applications on \amd{}.
We use native execution as the baseline and include native VMs and SGX enclaves as reference points.

\myparagraph{Applications}
We evaluate Lighttpd~\cite{lighttpd}, Hyper~\cite{rust-hyper}, SQLite~\cite{sqlite}, Redis~\cite{redis}, and llama-cpp~\cite{llama-cpp}.
Lighttpd and SQLite are common enclave workloads~\cite{graphene, veil} and appear in nested enclave studies such as Veil~\cite{veil}, enabling direct comparison.
Hyper is a high-performance, multi-threaded Rust HTTP server that scales with worker threads rather than processes~\cite{nginx}, avoiding \graminetyche{}’s unsupported fork.
HTTP servers are benchmarked with wrk~\cite{wrk}, Redis with memtier, SQLite with speedtest1~\cite{sqlite} at size \speedtestsize{} (which includes \speedtestinserts{} inserts), and LLaMa reports throughput~\cite{llama-cpp} after generating \llamanbtokens{} tokens.

\myparagraph{Configuration}
Binaries are from Gramine's default examples except for LLaMa (\autoref{sec:eval:use-case}).
Gramine applications use a \emph{manifest} specifying the number of threads and enclave memory.
We use default manifests~\cite{manifests} with \emph{exitless}~\cite{gramine-exitless} disabled due to \cveswitchlessone{} and \cveswitchlesstwo{}.
All enclaves include at least one worker and three Gramine runtime threads~\cite{gramine-threads} and 256MiB to 4GiB total memory.
For \system{}, we supply one physical core per manifest thread and configure native execution with the same worker-thread count.
Gramine-SGX and \graminetyche{} use the same application binaries, which we compiled from source without modification.

\myparagraph{\td{0}}
\td{0} shows no overhead in any benchmark compared to native, indicating that \system{} does not impact the performance of Linux applications.
This result is expected, as there are no calls or exits to the monitor and \td{0} has direct access to devices and the APIC.

\myparagraph{VMs \& CVMs}
\td{1} VMs match the performance of native VMs, demonstrating that \kvmdriver{} can support existing VM deployments.
For I/O benchmarks (\eg Hyper and Redis), \td{1} CVMs experience additional overheads due to extra copies through bounce buffers for I/O (see \autoref{sec:eval:overheads}).
They otherwise perform similarly to native VMs.%

\myparagraph{Enclaves}
\td{1} enclaves outperform SGX enclaves across all applications and are close to native for non-latency sensitive applications.
SGX performs poorly on LLaMa, even compared to an \td{2} enclave.
This is likely due to the limited EPC (\epcsize{}) size in SGX v1 compared to the high memory usage (4GiB) for the model~\cite{llama-meta, llama-model} and llama-cpp contexts~\cite{llama-cpp} and possibly memory encryption and integrity protection, thus we would expect SGX v2 to perform better.
Regardless, \td{2} only incurs a reasonable overhead compared to our baseline native execution as long as extra memory copies can be amortized.

\myparagraph{Memory usage}
\system{} requires private memory to store capabilities.
Running these workloads, we measured an average of two 4KiB pages per \td{} used for storing capabilities.
This extrapolates to, \eg a 4MiB memory overhead for 512 \td{}s.

\myparagraph{Discussion}
The latency of network operations is the primary issue but not a direct consequence of \system{}'s design, as discussed in \autoref{sec:eval:overheads}.
This is a known problem with confidential environments~\cite{teeio, bifrost, eleos, siopmp} that could be alleviated in \system{}, \eg with safe device passthrough.
\system{} otherwise demonstrates that existing frameworks can be ported to run \td{}s, achieving performance comparable to equivalent native deployments and outperforming Intel SGX, even in nested cases, reinforcing our claim of backward compatibility.

\begin{figure}[t]
    \includegraphics[width=\columnwidth]{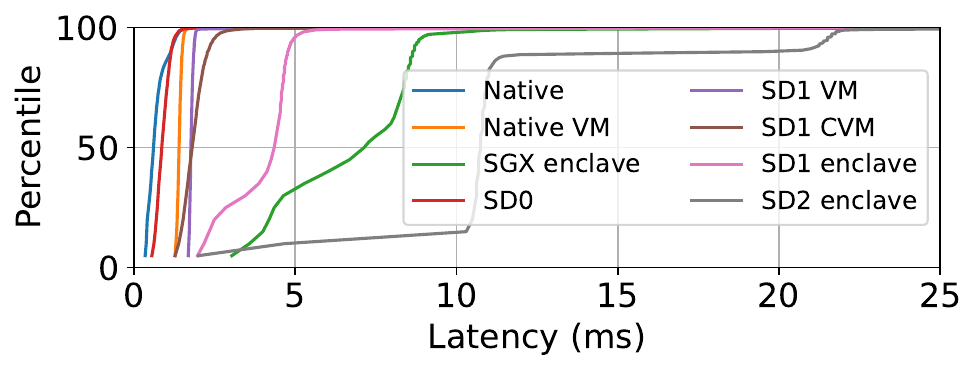}
    \caption{Redis GET latency distribution as measured by \textsf{memtier} (max throughput) during \autoref{fig:perf_comparison}'s experiment.}
    \label{fig:redis_latency}
\end{figure}

\subsection{Case Studies}
\label{sec:eval:use-case}

\myparagraph{Private inference with controlled sharing}
This case study uses \autoref{fig:llm-overview}'s scenario to protect user prompts from the CSP and model owner, while keeping model weights secure from the user and CSP.
The CSP runs a Linux \td{0} hypervisor with \kvmdriver{}; the user runs an \td{1} CVM that creates an \td{2} enclave for CPU-based LLM inference using llama-cpp~\cite{llama-cpp} with \graminetyche{} and four threads.
The ``proprietary model'' is an encrypted \llamamodel{}~\cite{llama-meta, llama-model}, decrypted inside the enclave as it is read from disk.
The enclave has no file or network access and shares memory only with the user’s \td{1} CVM, which receives prompts over SSH and forwards them to the enclave.
User keys can also be stored in a separate enclave.
As shown in \autoref{fig:perf_comparison}, the \td{2} enclave runs near native speed (\perfllama{} overhead), is ~10× faster than SGX enclaves, and goes beyond prior industrial solutions~\cite{apple-cloud-compute, edgeless} by also protecting the model from the user.
Future work on \kvmdriver{} may explore secure GPU passthrough.

\myparagraph{Managing trust within user-applications}
\system{}'s \td{}s are orthogonal to privilege levels and can compose isolation even within user-level code.
We prototyped compartmentalization of code within a user \td{} enclave.
While we did not provide a full port of an intra-process isolation framework, such a port could be done in the future; this prototype demonstrates the concept and fine-grained control over application TCB.

In this experiment, a user enclave running bearSSL~\cite{bearssl} hosts Redis inside a nested child \td{}, implemented either as a sandbox or a nested enclave. Both \td{}s minimize their TCB by including only a modified musl~\cite{musl-libc} libc, removing unnecessary system calls, and communicating through shared-memory queues. Before decrypting and forwarding requests to Redis, the SSL enclave attests that it fully encapsulates the Redis \td{}. Responses are returned via the shared queue and encrypted before being sent back.

A remote client (memtier) issues requests over a TLS connection with the SSL enclave. This experiment illustrates \system{}'s generality, showing that arbitrary composition and nesting of isolation boundaries is possible, even entirely within userspace.

\myparagraph{Sharing without the cost}
In existing cloud abstractions, communication between trusted TEEs -- enclaves or CVMs -- must bounce through untrusted memory, incurring software encryption, replay protection, integrity checks, and data copies to defend against tampering by untrusted software.
\system{} instead enables \td{}s to establish attestably private shared memory, accessible only to the two endpoints, eliminating this entire overhead.
This simplifies communication, improves performance, and enables confidential service VMs---trusted VMs that expose devices or services directly to other confidential endpoints.

We compare direct \td{}–to-\td{} communication with the traditional integrity-protect, encrypt, bounce, decrypt path.
Using AES-GCM and SHA-256 from the standard Linux crypto library, we measure added latencies of 20\microsecond for 1KB messages, 2ms for 1MB, and 0.24s for 100MB.
With \system{}, these costs are avoided entirely.
Physical attacks are orthogonal and can be handled in hardware, \eg with single- or multi-key total memory encryption (MK-TME)~\cite{mktme}, as discussed in \autoref{sec:threat-model}.

\section{Related Work}

\label{sec:related-work}

\system{} combines a security monitor design with an API inspired by micro-kernel to deliver a composable isolation solution.

\myparagraph{Security monitors}
They are trusted intermediaries~\cite{sys-design-principles} that enhance systems with new isolation and security guarantees~\cite{cloudvisor, overshadow, komodo} and are easier to inspect, update and extend than hardware solutions~\cite{hwissf}.
\arm{} CCA uses a monitor to support confidential VMs~\cite{cca} and Komodo~\cite{komodo} enables enclaves in \arm{} TrustZone~\cite{trustzone}.
Security monitors are popular in confidential computing to restrict privileged software, often through the use of virtualization, \eg
Inktag~\cite{inktag}, Overshadow~\cite{overshadow}, and HyperEnclave~\cite{hyperenclave} protect enclaves from untrusted OSes, Cloudvisor~\cite{cloudvisor} isolates VMs from hypervisors, and Blackbox~\cite{blackbox} secures containers.
They can offer intra-VM isolation, \eg Veil~\cite{veil} implements enclaves in AMD CVMs~\cite{sev} while Erebor~\cite{erebor} focuses on sandboxes in Intel CVMs~\cite{tdx}, or can harden kernel integrity~\cite{hypervision,nestedkernel,secvisor}.

Hardware vendors incrementally added privilege hierarchies within TEEs that can host such monitors -- Intel TDX~1.5~\cite{tdx-partitioning} introduces L1/L2 partitions, allowing a trusted L1~VMM to enforce intra-isolation in a TEE for up to three L2~VMs.
However, this remains subject to the limitations of intra-isolation discussed in this paper: the mechanism is platform-specific, hardware-capped in the number of sub-components, and scoped to a single CVM.

\system{} adopts a similar monitor-based implementation but differs in scope and abstraction.
It provides a unified isolation abstraction, \td{}s, that subsumes and composes the isolation boundaries traditionally enforced by separate security monitors, and applies to \textit{all} software on the machine -- not only components within a CVM but also across independently deployed VMs.
This design is inspired by micro-kernels.

\myparagraph{Micro-kernels}
Micro-kernels~\cite{minix, l4, sel4} follow a minimalist design, including only essential functions that cannot be implemented in user space, such as virtual memory, IPC, and thread management.
This reduces kernel complexity, improving security and maintainability.
By separating mechanisms from policies~\cite{lampson}, micro-kernels provide flexibility in resource management and isolation.
For instance, while the kernel handles page tables, user-space components configure virtual address spaces, enabling the \textit{recursive construction of address spaces}~\cite{l4} and custom isolation abstractions

\system{} leverages key micro-kernel concepts to deliver modern isolation guarantees.
Its memory operations resemble L4~\cite{l4}'s \emph{grant}, \emph{map}, and \emph{flush}.
They however differ in that \system{}'s operations focus on \textbf{attestable} isolation through guarantees on resource management: distinguishing explicitly between shared and exclusive resources and attesting memory is measured when received or scrubbed upon revocation (\autoref{sec:capabilities}).
\system{} also draws inspiration from Fluke~\cite{fluke}, which enables recursive VM isolation through ``nested processes'' without the overhead of naive nested virtualization.
\system{} generalizes this approach by providing a foundational mechanism to unify compartmentalization and compose confidential computing and encapsulation.

Despite its similarities to micro-kernel APIs, \system{} is \textit{not} a kernel replacement, does not create directly usable system abstractions such as processes, and requires an untrusted kernel in \td{0} to drive the machine.
The principles behind \system{}'s attestable isolation could however be backported into existing security-oriented micro-kernels~\cite{sel4}.

\myparagraph{Virtualization}
\system{} is not a hypervisor even though it runs in root-mode and uses virtualization extensions on \amd{}.
It does \textit{not} virtualize resources, provide full machine abstraction, or take allocation or scheduling decisions.
Instead, \system{} is closer to an exokernel~\cite{exokernel1} or a trusted state machine:
it validates operations against policies encoded by capabilities to correctly enforce and attest the partitioning and sharing of resources across \td{}s.
\system{} shares some similarities with hypervisors that adopt micro-kernel principles to improve security~\cite{hypsec, nova}, but these efforts focus on minimizing the hypervisor's TCB rather than providing a unified isolation abstraction to compose security boundaries.
Unlike type-1 hypervisors~\cite{hyperv, xen-v1}, which grant extra privileges to their initial domain (e.g., \texttt{DOM0} in Xen~\cite{xen-v1} or the root partition in Hyper-V~\cite{hyperv}), \td{0} in \system{} has no special privileges.

\myparagraph{Other mechanisms \& platforms}
Our \riscv{} prototype in \mmode{} demonstrates that \system{} does not require virtualization and can adapt to simpler access control mechanisms~\cite{nohype, soft-def-cpu, cheri, cheriot}.
\system{} could use alternative architectures~\cite{soft-def-cpu, nohype}, like NoHype~\cite{nohype}, that eliminate virtualization and provide simpler primitives to partition I/O, memory, and cores.
CHERI~\cite{cheri} capabilities offer a promising alternative to page-based mechanisms, already supporting enclaves~\cite{cheri-tree} and secure embedded devices~\cite{cheriot}.
\system{}'s software-defined capabilities provide a global view of system resources with extensible policies, complementing CHERI's efficient hardware enforcement for fine-grained memory isolation.

While we did not port \system{} to \arm{}, a virtualization-based backend could be implemented, running the monitor in EL2, similar to Blackbox~\cite{blackbox}, in normal, secure~\cite{cca}, or realm world~\cite{cca}.
Alternatively, it could run as firmware in EL3, akin to the \riscv{} backend, and leverage Granule Protection Table~\cite{cca} to isolate memory.
Similarly, \system{} could conceptually run as an Intel TDX~\cite{tdx} module, providing attestable policies for resource management and private shared memory across CVMs.
In practice, however, Intel is unlikely to allow custom TDX module implementations.
A more realistic alternative would be to deploy \system{} as an L1~VMM within a TDX CVM to isolate \td{}s running as L2s. This would bring clear semantics for isolation, controlled sharing within a CVM, and multi-component attestation, but would be limited to three L2s and would remain less general than our current approach, which provides these benefits at the scale of the entire machine rather than a single CVM and is portable across hardware platforms.

\balance

\section{Conclusion}

\system{} shows that trust management can become a first-class cloud abstraction.
Its security domain abstraction unifies composable isolation across privilege layers and system abstractions, shifting the burden of implementing and enforcing fine-grained isolation away from tenants.
We showed that \system{} is both general, by supporting and composing enclaves, sandboxes, and confidential virtual machines, and practical by running unmodified applications with near-native performance.
A confidential-inference case study with mutually distrustful users, model owners, and CSPs demonstrated its applicability to modern cloud workloads with complex trust models.

\section*{Acknowledgements}

We thank the many anonymous reviewers and shepherd,  Nate Foster, James R. Larus and Matthias Payer for their valuable feedback.
This work has received funding from the Swiss State Secretariat for Education, Research, and Innovation (SERI) under the SwissChips initiative, from the Microsoft-EPFL Joint Research Center, and gifts from the VMware University Research Fund.

\newpage

\bibliographystyle{plain}
\bibliography{gen-abbrev,dblp,misc}

\end{document}